\documentstyle[12pt]{article}
\begin{document}
\def\beq{\begin{equation}}
\def\eeq{\end{equation}}
\def\bea{\begin{eqnarray}}
\def\eea{\end{eqnarray}}
\def\ve{\vert}
\def\vel{\left|}
\def\ver{\right|}
\def\nnb{\nonumber}
\def\ga{\left(}
\def\dr{\right)}
\def\aga{\left\{}
\def\adr{\right\}}
\def\rar{\rightarrow}
\def\nnb{\nonumber}
\def\la{\langle}
\def\ra{\rangle}
\def\lla{\left<}
\def\rra{\right>}
\def\ba{\begin{array}}
\def\ea{\end{array}}
\def\tep{$B \rar K \ell^+ \ell^-$}
\def\tepm{$B \rar K \mu^+ \mu^-$}
\def\tept{$B \rar K \tau^+ \tau^-$}
\def\ds{\displaystyle}



\newskip\humongous \humongous=0pt plus 1000pt minus 1000pt
\def\caja{\mathsurround=0pt}
\def\eqalign#1{\,\vcenter{\openup1\jot
\caja   \ialign{\strut \hfil$\displaystyle{##}$&$
\displaystyle{{}##}$\hfil\crcr#1\crcr}}\,}


\def\simlt{\stackrel{<}{{}_\sim}}
\def\simgt{\stackrel{>}{{}_\sim}}



\def\bos{\lower 0.5cm\hbox{{\vrule width 0pt height 1.2cm}}}
\def\boss{\lower 0.35cm\hbox{{\vrule width 0pt height 1.cm}}}
\def\aaa{\lower 0.cm\hbox{{\vrule width 0pt height .7cm}}}
\def\dol{\lower 0.4cm\hbox{{\vrule width 0pt height .5cm}}}


\title{ {\Large {\bf 
Rare $B \rar K^\ast \nu \bar \nu$ decay beyond standard model } } }

\author{\vspace{1cm}\\
{\small T. M. Aliev \thanks
{e-mail: taliev@metu.edu.tr}\,\,,
A. \"{O}zpineci \thanks
{e-mail: ozpineci@newton.physics.metu.edu.tr}\,\,,
M. Savc{\i} \thanks
{e-mail: savci@metu.edu.tr}} \\
{\small Physics Department, Middle East Technical University} \\
{\small 06531 Ankara, Turkey} }
\date{}

\begin{titlepage}
\maketitle
\thispagestyle{empty}

\begin{abstract}
Using the most general, model--independent form of effective Hamiltonian,
the exclusive, rare $B \rar K^\ast \nu \bar \nu$ decay is analyzed. The
sensitivity of the branching ratios and missing mass--squared spectrum to
the new Wilson coefficients is discussed.
\end{abstract}

~~~PACS number(s): 12.60.--i, 13.20.He
\end{titlepage}

\section{Introduction}
Started operating, two $B$--factories BaBar and Belle \cite{R1}
open an excited era for studying $B$ meson physics and its rare decays. The main
physics program of these factories constitutes a detailed study of CP
violation in $B_d$ meson and precise measurement of rare
flavor--changing neutral current (FCNC) processes. The rare decays of 
$B$ mesons take
place via FCNC and appear in Standard Model (SM) at loop level. 
For this reason there appears a real possibility for checking the 
gauge structure of SM at loop level. 
On the other side rare decays are very sensitive to the new physics beyond
SM and their study is hoped to shed light on the existence of new particles 
before they are produced at colliders.

As has already been mentioned, one main goal of the $B$ physics program is
to find inconsistencies within the SM, in particular to find indications for  
new physics in the flavor and CP violating sectors \cite{R2}. 
In general, new physics effects manifest
themselves in rare $B$
meson decays either through new contributions to the
Wilson coefficients that exist in the SM or by introducing 
new structures in the effective Hamiltonian which are absent in the SM (see
\cite{R3}--\cite{R12}). Moreover, one can add new CP--violating phases and
modify the flavor changing neutral current. 

Currently the main interest is focused on the rare $B$ meson decays, for
which SM predicts "large" branching ratios. The rare 
$B \rar K^\ast \nu \bar \nu$ decay is such a decay that plays a special
role, both from experimental and theoretical point of view.

At quark level the FCNC $b \rar s \nu \bar \nu$ decay is described in
framework of the SM by the effective Hamiltonian \cite{R13}
\bea
\label{effH}
{\cal H}_{eff} &=& \frac{G_F\alpha}{2\sqrt{2} \pi \sin^2\theta_W}
V_{tb}V_{ts}^\ast X(x) \bar b \gamma^\mu (1-\gamma_5) s \bar \nu
\gamma_\mu (1-\gamma_5) \nu~,
\eea
where $G_F$ is the Fermi coupling constant, $\alpha$ is the fine structure
coupling constant, $\theta_W$ is the Weinberg angle, $V_{ij}$ is the
Cabibbo--Kobayashi--Maskawa (CKM) matrix elements and 
\bea
\label{Xx}
X(x) = X_0(x) + \frac{\alpha_s}{4 \pi} X_1(x)~,
\eea
where
\bea
\label{X0x}
X_0 = \frac{x}{8} \Bigg[\frac{x+2}{x-1} + \frac{3(x-2)}{(x-1)^2} \ln x
\Bigg]~,
\eea
and $x=m_t^2/m_W^2$.  
Explicit form of $X_0(x)$ and $X_1(x)$ can be found in \cite{R14} and 
\cite{R13}, respectively. Note that $X_1(x)$ gives about 3\%
contribution to the $X_0(x)$ term. The main attractive property of
(\ref{effH}) is that the $b \rar s \nu \bar \nu$ decay is governed only by a
single operator and is free of long distance effects related the presence of
four quark operators in the effective Hamiltonian (see for example
\cite{R15}). It is well known that the
situation for the $b \rar s \ell^+ \ell^-$ is more problematic, since this
decay is described by several Wilson coefficients, each bringing along its
own uncertainty, and also in this decay the long distance effects are essential 
and should be considered. In spite of all these theoretical advantages, it
might be very difficult to measure the inclusive mode $B \rar X_s \nu \bar
\nu$, because it requires to construct all $X_s$. Therefore it might be much
easier to measure the exclusive mode $B \rar K^\ast \nu \bar \nu$
experimentally. which has "large" branching ratio in SM, about $10^{-5}$
\cite{R16}.  

In this paper we study the $B \rar K^\ast \nu \bar \nu$ decay for a general
model independent form of effective Hamiltonian. 
$B \rar K^\ast \nu \bar \nu$ decay has  been extensively investigated in
framework of the SM and its various minimal 
extensions \cite{R16,R17}. 
Note that this mode was
studied in \cite{R18} in a similar way to our analysis, 
but in that work scalar and tensor
type interactions were introduced via  violation of the lepton number since 
neutrino was assumed to be
massless. But Super Kamiokande \cite{R19} results indicated that neutrino 
has mass.
Therefore neutrino has the right components and we can introduce scalar and
tensor interactions without any lepton number violation.
Another novel property of the present work is the appearence of new structures, e.g.,
terms that are proportional to $C_{LR}^{tot}$ and $C_{RR}$ (see Eq.
(\ref{matel}) below), which are absent in \cite{R18}.  

The paper is organized as follows. In section 2 we give the most general,
model independent form of effective Hamiltonian. We then parametrize the
long distance effects by appropriate form factors and calculate the
differential decay width. Section 3 is devoted to the numerical analysis and
concluding remarks.

\section{Differential decay width}
The exclusive $B \rar K^\ast \nu \bar \nu$ decay at quark level is described
by $b \rar s \nu \bar \nu$ transition. The decay amplitude for the $b \rar
s \nu \bar \nu$ decay in a general model independent form can be written as (for
general form of matrix element for the $b \rar s \ell^+ \ell^-$ decay, see
also \cite{R20,R21})
\bea
\label{matel}
{\cal M} &=& \frac{G_F\alpha}{4\sqrt{2} \pi \sin^2\theta_W}
V_{tb}V_{ts}^\ast \nnb \\
&\times&\Bigg\{
C_{LL}^{tot}\, \bar s \gamma_\mu (1-\gamma_5)b \,\bar \nu \gamma^\mu (1-\gamma_5)\nu +
C_{LR}^{tot} \,\bar s \gamma_\mu (1-\gamma_5)b \, \bar \nu \gamma^\mu (1+\gamma_5) \nu \nnb \\
&+&C_{RL} \,\bar s \gamma_\mu (1+\gamma_5) b \,\bar \nu \gamma^\mu (1-\gamma_5)\nu
+ C_{RR} \,\bar s \gamma_\mu (1+\gamma_5) b \, \bar \nu \gamma^\mu (1+\gamma_5) \nu \nnb \\
&+&C_{LRLR} \, \bar s (1+\gamma_5) b \,\bar \nu (1+\gamma_5) \nu
+C_{RLLR} \,\bar s (1-\gamma_5) b \,\bar \nu (1+\gamma_5)\nu \nnb \\
&+& C_{LRRL} \,\bar s (1+\gamma_5) b \,\bar \nu (1-\gamma_5) \nu +
C_{RLRL} \,\bar s (1-\gamma_5) b \,\bar \nu (1-\gamma_5) \nu \nnb \\
&+&C_T\, \bar s \sigma_{\mu\nu} b \,\bar \nu \sigma^{\mu\nu}\nu
+i C_{TE}\,\epsilon^{\mu\nu\alpha\beta} \bar s \sigma_{\mu\nu} b \,
\bar \nu \sigma_{\alpha\beta} \nu  \Bigg\}~.
\eea
Two of the four vector interactions containing $C_{LL}^{tot}$ and
$C_{LR}^{tot}$ already exist in the SM in combinations 
$(C_9-C_{10})$ and $(C_9+C_{10})$ for the $b \rar s \ell^+ \ell^-$ decay,
while in the present work for the $b \rar s \nu \bar \nu$ transition we have
$C_9=-C_{10}$. Therefore writing
\bea
C_{LL}^{tot} &=& 2 X + C_{LL}~, \nnb \\
C_{LR}^{tot} &=& C_{LR}~, \nnb
\eea 
one concludes that $C_{LL}^{tot}$ describes the sum of the contributions
from the SM and the new physics. The remaining coefficients 
$C_{LRLR},~C_{LRRL},~C_{RLLR},~C_{RLRL}$ and $C_T,~C_{TE}$ describe the
scalar and tensor interactions, respectively, which are absent in the SM.
The decay amplitude (\ref{matel}) has a rather general form as it includes nine
additional operators not found in the SM.

The decay amplitude of the semileptonic $B \rar K^\ast \nu \bar \nu$ decay
can be obtained after evaluating matrix elements of the quark operators in
Eq. (\ref{matel}) between the initial $\vel B(p_B) \rra$ and final 
$\lla K^\ast(p_{K^\ast},\varepsilon) \ver$ states. It follows 
from Eq. (\ref{matel}) that we need the following matrix elements 
\bea
\label{roll}
&&\lla K^\ast\vel \bar s \gamma_\mu (1 \pm \gamma_5)             
b \ver B \rra~,\nnb \\
&&\lla K^\ast \vel \bar s (1 \pm \gamma_5) b
\ver B \rra~, \nnb \\
&&\lla K^\ast \vel \bar s \sigma_{\mu\nu} b
\ver B \rra~. \nnb
\eea
These matrix elements can be written in terms of the form factors in the
following way:
\bea
\lefteqn{
\label{ilk}
\lla K^\ast(p_{K^\ast},\varepsilon) \vel \bar s \gamma_\mu 
(1 \pm \gamma_5) b \ver B(p_B) \rra =} \nnb \\
&&- \epsilon_{\mu\nu\lambda\sigma} \varepsilon^{\ast\nu} p_{K^\ast}^\lambda q^\sigma
\frac{2 V(q^2)}{m_B+m_{K^\ast}} \pm i \varepsilon_\mu^\ast (m_B+m_{K^\ast})   
A_1(q^2) \\
&&\mp i (p_B + p_{K^\ast})_\mu (\varepsilon^\ast q)
\frac{A_2(q^2)}{m_B+m_{K^\ast}}
\mp i q_\mu \frac{2 m_{K^\ast}}{q^2} (\varepsilon^\ast q)
\left[A_3(q^2)-A_0(q^2)\right]~,  \nnb \\  \nnb \\
\lefteqn{
\label{ucc}
\lla K^\ast(p_{K^\ast},\varepsilon) \vel \bar s \sigma_{\mu\nu} 
 b \ver B(p_B) \rra =} \nnb \\
&&i \epsilon_{\mu\nu\lambda\sigma}  \Bigg\{ - 2 T_1(q^2)
{\varepsilon^\ast}^\lambda (p_B + p_{K^\ast})^\sigma +
\frac{2}{q^2} (m_B^2-m_{K^\ast}^2) \Big[ T_1(q^2) - T_2(q^2) \Big] {\varepsilon^\ast}^\lambda 
q^\sigma \\
&&- \frac{4}{q^2} \Bigg[ T_1(q^2) - T_2(q^2) - \frac{q^2}{m_B^2-m_{K^\ast}^2} 
T_3(q^2) \Bigg] (\varepsilon^\ast q) p_{K^\ast}^\lambda q^\sigma \Bigg\}~. \nnb 
\eea
where $\varepsilon$ is the polarization vector of $K^\ast$ meson and $q = p_B-p_{K^\ast}$ 
is the momentum transfer.
To ensure finiteness of (\ref{ilk}) at $q^2=0$, 
it is usually assumed that $A_3(q^2=0) = A_0(q^2=0)$ and $T_1(q^2=0) = T_2(q^2=0)$.
The matrix element $\lla K^\ast \vel \bar s (1 \pm \gamma_5 ) b \ver B \rra$
can be calculated by 
contracting both sides of Eq. (\ref{ilk}) with $q^\mu$ and using equation of
motion. Neglecting the mass of the strange quark we get
\bea
\label{uc}
\lla K^\ast(p_{K^\ast},\varepsilon) \vel \bar s (1 \pm \gamma_5) b \ver
B(p_B) \rra =
\frac{1}{m_b} \Big[ \mp 2i m_{K^\ast} (\varepsilon^\ast q)
A_0(q^2)\Big]~.
\eea
In deriving Eq. (\ref{uc}) we have used the following relation
between the form factors $A_1~,A_2$ and $A_3$ (see \cite{R22})
\bea
A_3(q^2) = \frac{1}{2 m_{K^\ast}} \Big[ (m_B+m_{K^\ast}) A_1(q^2) -
(m_B-m_{K^\ast}) A_2(q^2)\Big]~. \nnb 
\eea
Taking into account Eqs. (\ref{matel}--\ref{uc}), the matrix element of the 
$B \rar K^\ast \bar \nu \nu$ decay can be written as 
\bea
\lefteqn{
\label{had}
{\cal M}(B\rightarrow K^\ast \nu \bar\nu) =
\frac{G_F \alpha}{4 \sqrt{2}\pi \sin^2\theta_W} V_{tb} V_{ts}^\ast }\nnb \\
&&\times \Bigg\{
\bar \nu \gamma^\mu (1-\gamma_5) \nu \, \Big[
-2 A \epsilon_{\mu\nu\lambda\sigma} \varepsilon^{\ast\nu}
p_{K^\ast}^\lambda q^\sigma
 -i B_1 \varepsilon_\mu^\ast
+ i B_2 (\varepsilon^\ast q) (p_B+p_{K^\ast})_\mu
+ i B_3 (\varepsilon^\ast q) q_\mu  \Big] \nnb \\
&&+ \bar \nu \gamma^\mu (1+\gamma_5) \nu \, \Big[
-2 C \epsilon_{\mu\nu\lambda\sigma} \varepsilon^{\ast\nu}
p_{K^\ast}^\lambda q^\sigma
 -i D_1 \varepsilon_\mu^\ast    
+ i D_2 (\varepsilon^\ast q) (p_B+p_{K^\ast})_\mu
+ i D_3 (\varepsilon^\ast q) q_\mu  \Big] \nnb \\
&&+\bar \nu (1-\gamma_5) \nu \Big[ i B_4 (\varepsilon^\ast
q)\Big]
+ \bar \nu (1 + \gamma_5) \nu \Big[ i B_5 (\varepsilon^\ast
q)\Big]  \nnb \\
&&+4 \bar \nu \sigma^{\mu\nu}  \nu \Big( i C_T \epsilon_{\mu\nu\lambda\sigma}
\Big) \Big[ -2 T_1 {\varepsilon^\ast}^\lambda (p_B+p_{K^\ast})^\sigma +
B_6 {\varepsilon^\ast}^\lambda q^\sigma -
B_7 (\varepsilon^\ast q) {p_{K^\ast}}^\lambda q^\sigma \Big] \nnb \\
&&+16 C_{TE} \bar \nu \sigma_{\mu\nu}  \nu \Big[ -2 T_1
{\varepsilon^\ast}^\mu (p_B+p_{K^\ast})^\nu  +B_6 {\varepsilon^\ast}^\mu q^\nu -
B_7 (\varepsilon^\ast q) {p_{K^\ast}}^\mu q^\nu
\Bigg\}~,
\eea
where
\bea
\label{as}
A &=& (C_{LL}^{tot} + C_{RL}) \frac{V}{m_B+m_{K^\ast}}~, \nnb \\
B_1 &=& (C_{LL}^{tot} - C_{RL}) (m_B+m_{K^\ast}) A_1~, \nnb \\
B_2 &=& (C_{LL}^{tot} - C_{RL})
\frac{A_2}{m_B+m_{K^\ast}}~, \nnb \\
B_3 &=& 2 (C_{LL}^{tot} - C_{RL}) m_{K^\ast} \frac{A_3-A_0}{q^2}~, \nnb \\
C &=& A(C_{LL}^{tot} \rar C_{LR}^{tot},~C_{RL} \rar C_{RR}) ~,\nnb \\
D_1 &=& B_1(C_{LL}^{tot} \rar C_{LR}^{tot},~C_{RL} \rar C_{RR}) ~,\nnb \\
D_2 &=& B_2(C_{LL}^{tot} \rar C_{LR}^{tot},~C_{RL} \rar C_{RR}) ~,\nnb \\
D_3 &=& B_3(C_{LL}^{tot} \rar C_{LR}^{tot},~C_{RL} \rar C_{RR}) ~,\nnb \\
B_4 &=& - 2 ( C_{LRRL} - C_{RLRL}) 
\frac{ m_{K^\ast}}{m_b} A_0 ~,\nnb \\
B_5 &=& - 2 ( C_{LRLR} - C_{RLLR}) \frac{m_{K^\ast}}{m_b} A_0 ~,\nnb \\
B_6 &=& 2 (m_B^2-m_{K^\ast}^2) \frac{T_1-T_2}{q^2} ~,\nnb \\
B_7 &=& \frac{4}{q^2} \left( T_1-T_2 - 
\frac{q^2}{m_B^2-m_{K^\ast}^2} T_3 \right)~. \nnb    
\eea 
Note that in further calculations we set neutrino mass to zero.

From the matrix element (\ref{had}) it is straightforward to derive the
missing mass--squared spectrum corresponding to the longitudinally and
transversally polarized $K^\ast$ meson. In the case of longitudinally
polarized $K^\ast$ meson, we get for the missing mass--squared spectrum
\bea
\label{lon}
\frac{d \Gamma_L}{d q^2} &=& N_\nu 
\Bigg[\frac{G_F \alpha}{4 \sqrt{2}\pi \sin^2\theta_W}\Bigg]^2
\vel V_{tb} V_{ts}^\ast \ver^2
\frac{1}{256 m_B^2 \pi^3} \, \frac{1}{3 m_{K^\ast}^2} \nnb \\
&\times& \Bigg\{
4 \vel 2 B_1 h + B_2 \lambda \ver^2 + 
4 \vel 2 D_1 h + D_2 \lambda \ver^2 
+ 6 \Big( \vel B_4 \ver^2 + \vel B_5 \ver^2 \Big) \lambda q^2\nnb \\ 
&+& 16 \vel 4 B_6 h - B_7 \lambda \ver^2
\Big( 4 \vel C_{TE} \ver^2 + \vel C_T \ver^2 \Big) q^2 \times \Bigg[
16 \vel T_1 \ver^2 (m_B^2 + 3 m_{K^\ast}^2 - q^2 )^2 \nnb \\
&-& 16 \mbox{\rm Re} (B_6 T_1^\ast) 
\Big[\lambda + 4 m_{K^\ast}^2 (m_B^2 - m_{K^\ast}^2) \Big] +
8\mbox{\rm Re} (B_7 T_1^\ast) (\lambda + 3 m_{K^\ast}^2 - q^2) 
\Bigg] \Bigg\}~.
\eea

For the transversally polarized $K^\ast$ meson, the differential decay width
takes the following form
\bea
\label{trn}
\frac{d \Gamma_\mp}{d q^2} &=& N_\nu
\Bigg[\frac{G_F \alpha}{4 \sqrt{2}\pi \sin^2\theta_W}\Bigg]^2
\vel V_{tb} V_{ts}^\ast \ver^2 \nnb \\
&\times& \Bigg\{
\frac{16}{3} \sqrt{\lambda} \, q^2 \Big(
\vel A \pm B_1 \ver^2 + \vel C \pm D_1 \ver^2 \Big)\nnb \\
&+& \frac{2048}{3} \sqrt{\lambda} \, \mbox{\rm Re} (C_T C_{TE}^\ast)
\Big[ \mp 4 (m_B^2 - m_{K^\ast}^2) \vel T_1 \ver^2 + 
2 \mbox{\rm Re} (B_6 T_1^\ast) q^2 \Big] \nnb \\
&+& \frac{256}{3} \Big( 4 \vel C_{TE} \ver^2 + \vel C_T \ver^2 \Big) 
\Big( \vel B_6 \ver^2 q^4 + 4 \big[ \lambda + (m_B^2 - m_{K^\ast}^2) \big] 
\vel T_1 \ver^2 \nnb \\
&-& 4 (m_B^2 - m_{K^\ast}^2) \mbox{\rm Re} (B_6 T_1^\ast) q^2 \Big) \Bigg\} 
\eea 
In Eqs. (\ref{lon}) and (\ref{trn}) $N_\nu=3$ is the number of light neutrinos, 
\bea
\lambda(m_B^2,m_{K^\ast}^2,q^2) &=& 
m_B^4 + m_{K^\ast}^4 + q^4 - 2 m_B^2 q^2 - 2 m_{K^\ast}^2 q^2 - 
2 m_B^2 m_{K^\ast}^2 ~, \nnb \\
h &=& \frac{1}{2} (m_B^2 - m_{K^\ast}^2 - q^2) ~. \nnb 
\eea 

It should be noted that in experiments due to the non--detectibility of the
neutrinos, it is impossible to discriminate the transverse polarization $+1$
from $-1$. For this reason these two polarization states must be added,
i.e.,
\bea
\label{lontrn}
\frac{d \Gamma_T}{d q^2}(B \rar K^\ast \nu \bar \nu) = 
\frac{d \Gamma_+}{d q^2}(B \rar K^\ast \nu \bar \nu) +
\frac{d \Gamma_-}{d q^2}(B \rar K^\ast \nu \bar \nu)~.
\eea
From Eqs. (\ref{lon}) and (\ref{trn}) we observe that scalar interaction
gives contribution to the differential decay width $d\Gamma_L/dq^2$ 
when $K^\ast$ is longitudinally polarized and does not contribute to 
$d\Gamma_\pm/dq^2$ when it transversally polarized. This result can be
explained in the following way. When $B$ meson is at rest, $K^\ast$ meson and
neutrino pair must be in flight along opposite directions. When $K^\ast$
meson is transversally polarized the total helicity of the neutrino pair
must be $\pm 1$, since the initial $B$ meson spin is equal to zero. But from
the neutrino antineutrino pair, which are in flight along the same direction,
one can organize a total helicity of $\pm 1$ by flipping one of the
neutrino's helicity. But this spin flip can be achieved by inserting mass of
neutrino.
However, as has already been mentioned previously, we neglect
the neutrino mass in our calculations and for this reason in the expression
for the differential decay width when  $K^\ast$ meson is transversally
polarized, the terms describing the scalar interaction disappear. 

\section{Numerical analysis}    
In this section we will study the sensitivity of the branching ratio and
missing mass--squared spectrum to the new Wilson coefficients. 

In performing numerical calculations, as can easily be seen from Eqs.
(\ref{lon}) and (\ref{trn}), first of all, we need the expressions for the
form factors.  
For the values of the form factors, we have used the results of
\cite{R23} (see also \cite{R24} and \cite{R25}), 
where  the radiative corrections to the leading twist
contribution and $SU(3)$ breaking effects are also taken into account.
It is shown in \cite{R23} that  
the $q^2$ dependence of the form factors can be represented in terms of 
three parameters as
\bea
F(q^2) = \frac{F(0)}{\ds{1-a_F\,\frac{q^2}{m_B^2} + b_F \left
    ( \frac{q^2}{m_B^2} \right)^2}}~, \nnb
\eea
where, the values of parameters $F(0)$, $a_F$ and $b_F$ for the
$B \rar K^\ast$ decay are listed in Table 1.

\begin{table}[h]                    
\renewcommand{\arraystretch}{1.5}
\addtolength{\arraycolsep}{3pt}
$$
\begin{array}{|l|ccc|}
\hline
& F(0) & a_F & b_F \\ \hline
A_1^{B \rar K^*} &
\phantom{-}0.34 \pm 0.05 & 0.60 & -0.023 \\
A_2^{B \rar K^*} &
\phantom{-}0.28 \pm 0.04 & 1.18 & \phantom{-}0.281\\
V^{B \rar K^*} &
 \phantom{-}0.46 \pm 0.07 & 1.55 & \phantom{-}0.575\\
T_1^{B \rar K^*} &
  \phantom{-}0.19 \pm 0.03 & 1.59 & \phantom{-}0.615\\
T_2^{B \rar K^*} & 
 \phantom{-}0.19 \pm 0.03 & 0.49 & -0.241\\
T_3^{B \rar K^*} & 
 \phantom{-}0.13 \pm 0.02 & 1.20 & \phantom{-}0.098\\ \hline
\end{array}   
$$
\caption{$B$ meson decay form factors in a three-parameter fit, where the
radiative corrections to the leading twist contribution and SU(3) breaking  
effects are taken into account \cite{R25}.}
\renewcommand{\arraystretch}{1}
\addtolength{\arraycolsep}{-3pt}
\end{table}

In Figs. (1) and (2) we present the dependence of the branching ratios 
on the new Wilson coefficients,
when $K^\ast$ polarized transversally and longitudinally, respectively. 
From both figures we see that when $C_{LL}$ lies in the region from $-4$ to $0$
(in numerical calculations all new Wilson coefficients vary in the range
from $-4$ to $+4$), branching
ratios are lower than the SM prediction.
Moreover, when $C_{LL}$ increases from $0$ up to $+4$, branching
ratios become larger than the SM result and for increasing values of
$C_{LL}$ the departure from SM becomes substantial. This behavior can be
explained by the fact that in the range from $-4$ to $0$ the new Wilson
coefficient $C_{LL}$ gives destructive and in the second half of
the range from $0$ to $+4$ it gives constructive interference to the SM
result.
For the Wilson
coefficients $C_{RR}$ and $C_{LR}$, we observe the following dependence
of the branching ratios. 
Up to the zero
value of the Wilson coefficients the branching ratios decrease and at
$C_{LR} = C_{RR} =0$ they coincide with the SM prediction. Furthermore, with
increasing $C_{RR}$, $C_{LR}$ both ${\cal B}_L$ and ${\cal B}_T$ increase. 
Qualitatively, this behavior could be explained as follows. 
When all Wilson coefficients, except $C_{RR}$ (or $C_{LR}$), are zero,
${\cal B}_L$ and ${\cal B}_T$ are proportional to $\vel C_{RR} \ver^2$ 
(or $\vel C_{LR} \ver^2$). Therefore as $C_{RR}$ (or $C_{LR}$) increases in
the region from $-4$ to $0$, ${\cal B}_L$ and ${\cal B}_T$ decrease and when 
$C_{RR}$ (or $C_{LR}$) increase from $0$ to $+4$ the above--mentioned grow
larger. Obviously the dependence of ${\cal B}_L$ and ${\cal B}_T$ on 
$C_{RR}$ (or $C_{LR}$) must be symmetric, and this expectation is confirmed
by the numerical calculations.  
In the case
of the dependence of branching ratios on the Wilson coefficient $C_{RL}$, we
observe that ${\cal B}_L$ decreases with changing values of $C_{RL}$ 
in the range from $-4$ to $+4$. We can argue about this dependence as follows.
For this Wilson coefficient ${\cal B}_L$ is proportional to 
$\vel 2 X - C_{RL} \ver^2$, and hence ${\cal B}_L$ decreases for increasing
values of $C_{RL}$, as expected.  
However the situation is different for the ${\cal B}_T$, 
as can easily be seen from the figure, it decreases when $C_{RL}$ increases 
from $-4$ to $0$ and
then increases in the positive half of the range. 

As has already been noted, ${\cal B}_T$ is independent of the scalar type
interaction while ${\cal B}_L$ is dependent.   
From Fig. (2) we observe that for all scalar type interaction coefficients
the branching ratio ${\cal B}_L$ shows a similar behavior, i.e., it
decreases in the first half of the range of variation of the scalar
interaction coefficients and increases for the positive part of the range 
from $0$ to $+4$.
In contrary to the previous cases, as Figs. (1) and (2) depict, ${\cal B}_L$
and ${\cal B}_T$ show quite a strong dependence on the tensor type
interaction coefficients $C_T$ and $C_{TE}$. As can easily be seen from 
Eqs. (\ref{lon}) and (\ref{trn}), ${\cal B}_L$ and ${\cal B}_T$ depend as
moduli square on $C_T$ and $C_{TE}$ and therefore this dependence must be symmetric.
If we assume that the departure from SM prediction is expected to
be small, it will put very strong restriction to the tensor type
interactions.

We also analyze the missing mass--squared spectrum on new Wilson
coefficients. All qualitative arguments which we have put forward in
discussing the dependence of branching ratios on new Wilson coefficients 
work their way similarly and remains in power in the case of missing
mass--squared spectrum as well. As an example in Fig. (3) we present the
dependence of missing mass--squared spectrum at four different values of
$C_{LL}$, namely $-2,~-1~,0,~+1,~+2$.

Finally a few words about the dependence of the another experimentally
measurable quantity, namely ${\cal B}_L/{\cal B}_T$, on the new Wilson
coefficients are in order. 
Our numerical analysis shows that this ratio is practically independent of
the new Wilson coefficients and are very close to the SM prediction.
Therefore study of this ratio can not serve as an effective tool in search
of new physics beyond SM. 
  
In conclusion, using the most general, model independent form of the
effective Hamiltonian we have studied the sensitivity of the branching
ratios ${\cal B}_L$ and ${\cal B}_T$ to the new Wilson coefficients. 
The main result of this study is that the
branching ratios and the missing mass--squared spectrum are very 
useful in looking  new physics beyond SM.

\newpage
\section*{Figure captions}
{\bf Fig. (1)} The dependence of the branching ratio of the  $B \rar K^\ast
\nu \bar \nu$ decay on the new Wilson coefficients,
when $K^\ast$ polarized transversally. The line indicated by $C_{XXXX}$
denotes any one of the four scalar interaction coefficients, namely,
$C_{LRRL},~C_{RLLR},~C_{LRLR}$ and $C_{RLRL}$.\\ \\
{\bf Fig. (2)} The same as in Fig. (1), but when $K^\ast$ polarized
longitudinally. \\ \\
{\bf Fig. (3)} The dependence of missing mass--squared spectrum at four different values of
$C_{LL}$, namely $-2,~-1~,0,~+1,~+2$. The first four lines represent
the case when $K^\ast$ polarized longitudinally, and the remaining four lines 
represent the case  when $K^\ast$ polarized transversally.
\newpage

\begin{figure}
\vskip 1cm
    \includegraphics{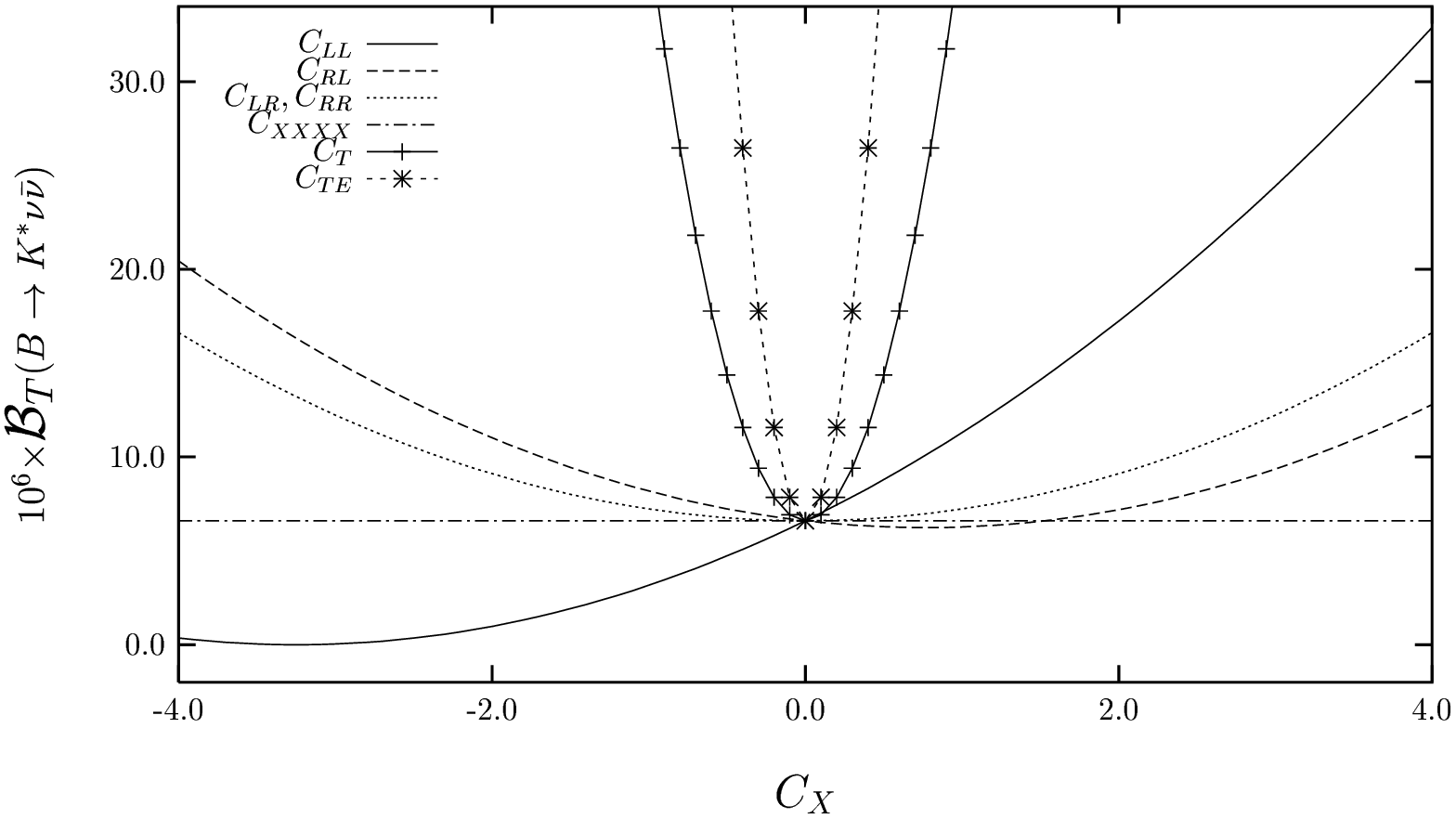}
\vskip 8.1cm
\caption{}
\end{figure}

\begin{figure}
\vskip 1.5 cm
    \includegraphics{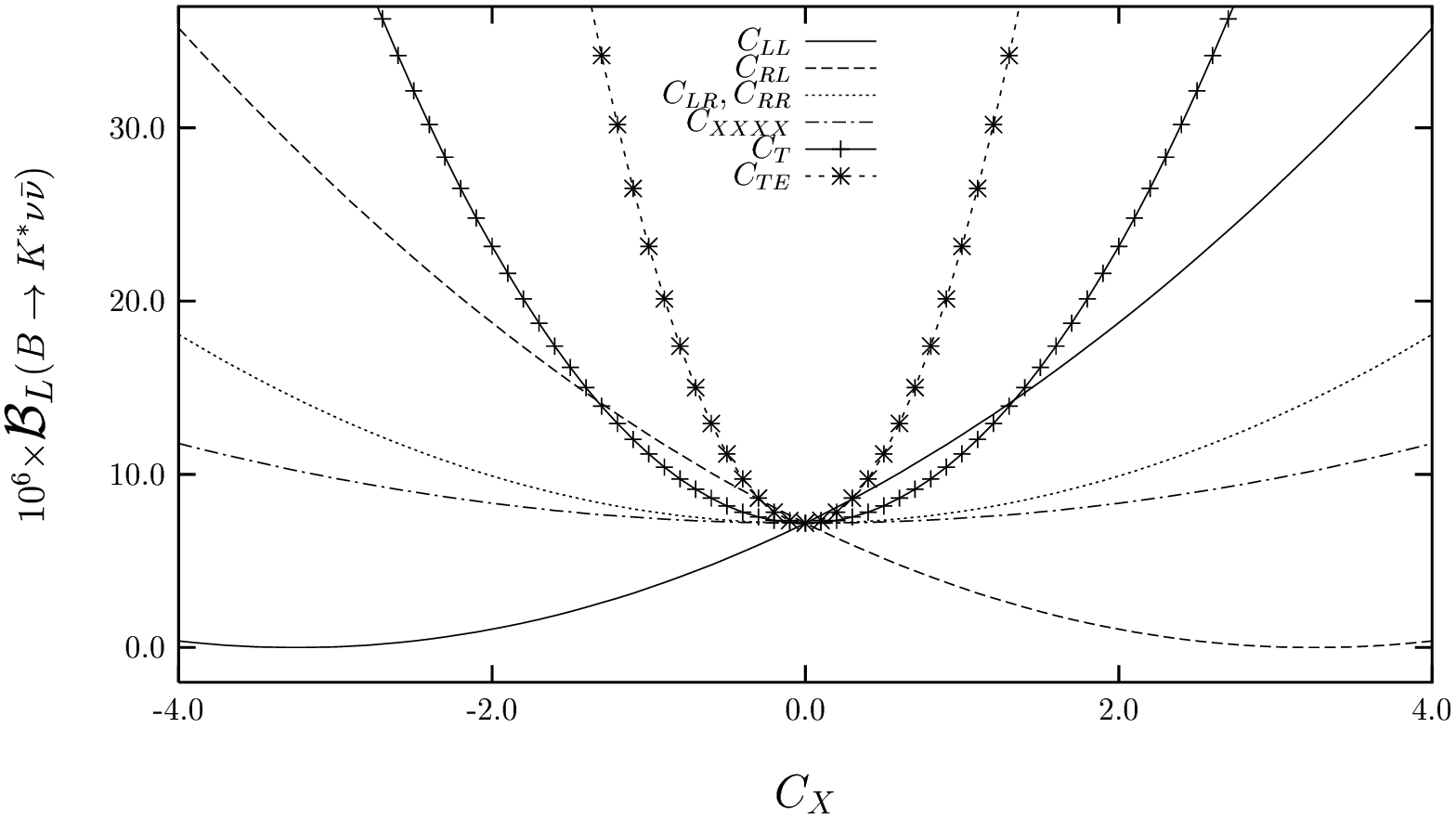}
\vskip 9. cm
\caption{}
\end{figure}

\begin{figure}
\vskip 1.5 cm
    \includegraphics{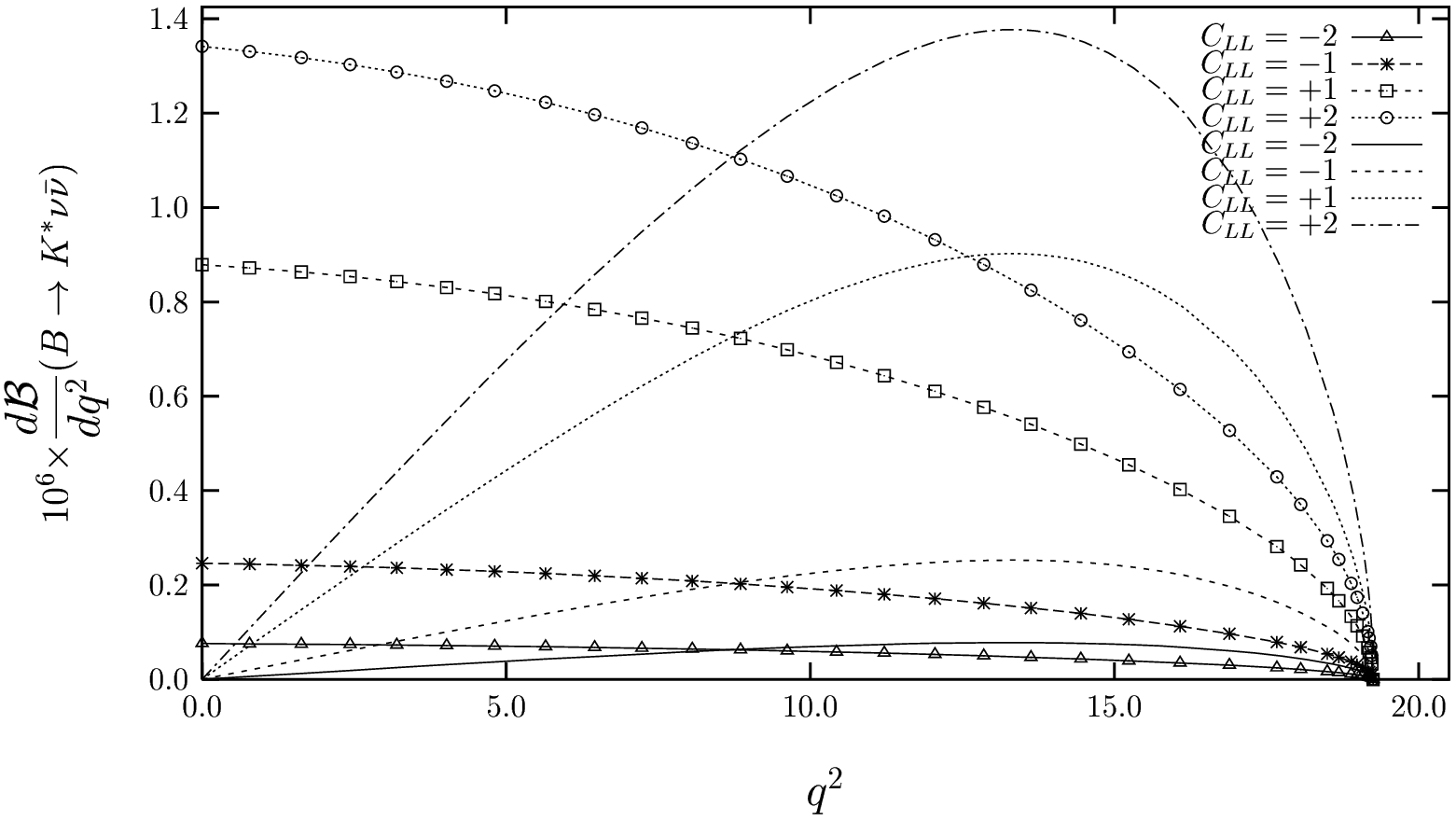}
\vskip 9. cm
\caption{}
\end{figure}

\end{document}